\documentclass[a4paper]{jpconf}
\usepackage{graphicx}
\begin{document}
\title{Entanglement and separability in the noncommutative phase-space scenario}

\author{Alex E. Bernardini}
\address{Departamento de F\'{\i}sica, Universidade Federal de S\~ao Carlos, PO Box 676, 13565-905, S\~ao Carlos, SP, Brasil.}
\ead{alexeb@ufscar.br}
\author{Catarina Bastos}
\address{Instituto de Plasmas e Fus\~ao Nuclear, Instituto Superior T\'ecnico Avenida Rovisco Pais 1, 1049-001 Lisboa, Portugal}
\ead{catarina.bastos@ist.utl.pt}
\author{Orfeu Bertolami}
\address{Departamento de F\'isica e Astronomia, Faculdade de Ci\^encias da Universidade do Porto, Rua do Campo Alegre, 687,4169-007 Porto, Portugal}
\ead{orfeu.bertolami@fc.up.pt}
\author{Nuno Costa Dias and Jo\~ao Nuno Prata \footnote[1]{Also at
Grupo de F\'{\i}sica Matem\'atica, UL, Avenida Prof. Gama Pinto 2, 1649-003, Lisboa, Portugal.}}
\address{Departamento de Matem\'{a}tica, Universidade Lus\'ofona de Humanidades e Tecnologias Avenida Campo Grande, 376, 1749-024 Lisboa, Portugal}
\ead{ncdias@meo.pt, joao.prata@mail.telepac.pt}
\begin{abstract}
Quantumness and separability criteria for continuous variable systems are discussed for the case of a noncommutative (NC) phase-space.
In particular, the quantum nature and the entanglement configuration of NC two-mode Gaussian states are examined.
Two families of covariance matrices describing standard quantum mechanics (QM) separable states are deformed into a NC QM configuration and then investigated through the positive partial transpose criterium for identifying quantum entanglement.
It is shown that the entanglement of Gaussian states may be exclusively induced by switching on the NC deformation.
Extensions of some preliminary results are presented.
\end{abstract}

\section{Introduction}
The most salient quantum features in the context of a phase-space noncommutative (NC) extension of quantum mechanics (QM) have already been quantified \cite{Bastos} and issues such as quantum entanglement \cite{Bernardini13B}, quantum beating, wave function collapse and loss of quantum coherence \cite{Bernardini13A} have also been examined.
These findings allow for a broader understanding of the correspondence between quantum and classical systems, as well as for probing the role of the NC deformation when tuning to standard QM predictions.

The NC extensions of QM show an impressive range of characteristic features when concerned on studies of the quantum Hall effect \cite{Prange}, on the the gravitational quantum well for ultra-cold neutrons \cite{06A}, on the Landau level and the $2D$ harmonic oscillator problems in the phase-space \cite{Nekrasov01,Rosenbaum}, and as a probe for quantum beating and missing information effects \cite{Bernardini13A} as well as for quantum entanglement issues \cite{Bernardini13B}.
This NC QM scenario also admits putative violations of the Robertson-Schr\"odinger uncertainty relation \cite{Catarina002}.
On a broader context, noncommutativity is believed to be relevant in the formalism of quantum gravity and string theory \cite{Connes,Douglas,Seiberg}.
Furthermore, in the framework of quantum cosmology, phase-space noncommutativity has shown to bring up novel features in what concerns the black hole singularity \cite{Bastos3,Bastos3B,Bastos3C,Bastos3D}, as well as to the equivalence principle \cite{Bastos4}.

From the theoretical perspective, the NC QM is supported by extensions of the Heisenberg-Weyl algebra \cite{Bastos,06A,08A,09A,Gamboa}.
The theory lives in a $2d$-dimensional phase-space where time is required to be a commutative parameter, and coordinate and momentum variables obey a NC algebra.
In particular, the NC extension of QM is more suitably formulated in the Weyl-Wigner-Groenewold-Moyal (WWGM) formalism for QM \cite{Groenewold,Moyal,Wigner}.
From the subjacent algebra perspective, the phase-space NC generalizations of QM are based on extensions of the Heisenberg-Weyl algebra \cite{Bastos,08A,GossonAdhikari}.
NC QM shows also to have a bearing on issues such as the Osawa's uncertainty principle (OUP)\cite{Osawa,Osawa3,RSUP2} as discussed in Ref.~\cite{RSUP1}. Relevantly, the OUP is claimed to be the one supported experimentally \cite{Stein,Hasegawa}.

The proposal of this contribution is to extend the formalism for describing quantumness and separability of two-mode Gaussian states, setting up a NC criterion for entanglement.
We report on some previous results on the phase-space NC QM \cite{Bernardini13B}, more specifically on the role of phase-space noncommutativity to yield Gaussian entangled states.
Concerning the framework for identifying the separability and the entanglement of quantum Gaussian states, one of the main results reported in this manuscript is related to the positive partial transposed (PPT) separability criterion \cite{Peres,Horo}. 
Generically, the PPT criterion provides a necessary and, in some cases, sufficient condition for distinguishing between separable and entangled states in discrete quantum systems.
The framework has been extended to continuous variable systems \cite{Simon} by implementing the partial transpose operation as a mirror reflection in the Wigner phase-space.
The continuous PPT criterion presented in the theory of quantum information of Gaussian states \cite{gauss01,Adesso, Weedbrook} has been examined in the context of identifying the entanglement properties exclusively induced by a NC deformation of phase-space \cite{Bernardini13B}.
Besides recovering the results of Ref.~\cite{Bernardini13B}, hereon novel results for typical Gaussian states are presented.

\section{Noncommutative criterion for entanglement and separability}

Consider a bipartite quantum system described in terms of a $2n_A$-dimensional subsystem $A$ and a $2n_B$-dimensional subsystem $B$ with $n_A+n_B=n$, one can write the collective degrees of freedom of the composite quantum system as
$\widehat{z} = (\widehat{z}^A,\,\widehat{z}^B)$, where $\widehat{z}^A =(\widehat{x}_1^A,
\cdots,\widehat{x}_{n_A}^A,\, \widehat{p}_1^A, \cdots, \widehat{p}_{n_A}^A)$ and $\widehat{z}^B
=(\widehat{x}_1^B, \cdots,\widehat{x}_{n_B}^B,\, \widehat{p}_1^B, \cdots, \widehat{p}_{n_B}^B)$, which correspond to the generalized coordinates of the two subsystems \cite{Bernardini13B}.
The assumed commutation relations given by
\begin{equation}
\left[\widehat{z}_i, \widehat{z}_j \right] = i \Omega_{ij}, \hspace{1 cm} i,j = 1, \cdots, 2n,
\label{eq1}
\end{equation}
are described in terms of the associated matrix given by ${\bf \Omega} = \left[\Omega_{ij} \right] \equiv Diag\left[\bf \Omega^A,\,\bf \Omega^B\right]$,
where ${\bf \Omega^K}$, with $K = A$ and $B$, is a real skew-symmetric non-singular $2n_K \times 2 n_K$ matrix of the form
\begin{equation}
{\bf \Omega^K} = \left(
\begin{array}{c c}
{\bf \Theta^K} & {\bf I^K}\\
- {\bf I^K} & {\bf \Upsilon^K}
\end{array}
\right),
\label{eq2}
\end{equation}
where ${\bf \Theta^K} = \left[\theta_{ij}^K \right]$ and ${\bf \Upsilon^K} = \left[\eta_{ij}^K \right]$, respectively, measure the noncommutativity of the position-position and momentum-momentum sectors of the subsystems $A$ and $B$, with ${\bf I^K}$ as the $n_K \times n_K$ identity matrix (and it has been set $\hbar =1$).
The NC structure can be formulated in terms of commuting variables by considering a linear Darboux transformation (DT), $\widehat{z} = {\bf S} \widehat{\zeta}$, where $ \widehat{\zeta}=( \widehat{\zeta}^A,
\widehat{\zeta}^B)$, with $ \widehat{\zeta}^A= ( \widehat{q}_1^A, \cdots, \widehat{q}_{n_A}^A, \, \widehat{k}_1^A,
\cdots, \widehat{k}_{n_A}^A)$ and $ \widehat{\zeta}^B= ( \widehat{q}_1^B, \cdots, \widehat{q}_{n_B}^B,\,
\widehat{k}_1^B, \cdots, \widehat{k}_{n_B}^B)$ satisfying the usual QM commutation relations:
\begin{equation}
\left[\widehat{\zeta}_i, \widehat{\zeta}_j \right] = i J_{ij} , \hspace{1 cm} i,j=1, \cdots, 2n,
\label{eq3}
\end{equation}
where ${\bf J} = \left[J_{ij} \right]=Diag \left[{\bf J^A},{\bf J^B} \right] $ with
\begin{equation}
{\bf J^K} = - ({\bf J^K})^T = - ({\bf J^K})^{-1} = \left(
\begin{array}{c c}
{\bf 0} & {\bf I^K}\\
- {\bf I^K} & {\bf 0}
\end{array}
\right)~,
\label{eq4}
\end{equation}
which is a $2n_K \times 2n_K$ standard symplectic matrix for $K =
A$ and $B$. The linear transformation ${\bf S} \in Gl(2n)$ is such
that ${\bf S} = Diag\left[{\bf S^A},\,{\bf S^B}\right]$, and ${\bf
\Omega} ={\bf S} {\bf J} {\bf S}^T  $, or in terms of submatrices labelled by $K$, ${\bf
\Omega^K} ={\bf S^K} {\bf J^K} ({\bf S^K})^T$.
Giving that the map ${\bf S}$ is not uniquely defined, if one
composes ${\bf S}$ with block-diagonal canonical transformations, an equally valid DT is obtained.


For a composite system described by the density matrix $\rho$, function of the NC variables $\widehat{z}$, the DT yields the density matrix $\widetilde{\rho} (\widehat{\zeta}) = \rho \left({\bf S}
\widehat{\zeta} \right)$, which is associated with the Wigner function: \small \begin{equation} W \widetilde{\rho} (\zeta) = \! {1\over{(2 \pi)^{n}}} \int_{{R}^{n_A}}\!\! \!\!dy^A
\!\!\int_{{R}^{n_B}} \!\!\!\!dy^B e^{- i \left(y^A \cdot k^A + y^B \cdot k^B\right)} \left\langle q^{A} +{y^A  \over 2},\,q^{B} +{y^B  \over 2} \left\vert\, \widetilde{\rho}
\,\right\vert q^{A} -{y^A  \over 2},\,q^{B} -{y^B  \over 2}\right\rangle . \label{eq5} \end{equation} \normalsize Upon inversion of
the DT, one obtains the NC Wigner function
\cite{Bastos}:
\begin{equation}
W \rho (z) = (\det {\bf \Omega})^{-1/2} W \widetilde{\rho} ({\bf S}^{-1} z).
\label{eq6}
\end{equation}
If ${\bf \Sigma}$ denotes the covariance matrix of $W \rho$ and ${\bf \widetilde{\Sigma}}$ that of $W \widetilde{\rho}$, then the two are related by:
\begin{equation}
{\bf \Sigma} = {\bf S} {\bf \widetilde{\Sigma}} {\bf S}^T.
\label{eq7}
\end{equation}
A necessary condition for the phase-space function $W \widetilde{\rho}$ with covariance matrix ${\bf
\widetilde{\Sigma}}$ to be an admissible (commutative) Wigner function is that it satisfies the
Robertson-Schr\"odinger uncertainty principle (RSUP),
\begin{equation}
{\bf \widetilde{\Sigma}} + (i/2) {\bf J} \ge 0,
\label{eq8B}
\end{equation}
that is, a $2n \times 2n$ positive matrix in ${C}^{2n}$. From this condition and Eq.~(\ref{eq7}) one concludes that for $W \rho$ to be an equally admissible NC Wigner function, it has to satisfy the NC RSUP \cite{Catarina002}:
\begin{equation}
{\bf \Sigma} + (i/2) {\bf \Omega} \ge 0.
\label{eq8}
\end{equation}
For Gaussian states this condition is also sufficient \cite{Catarina002}.

A composite quantum system is separable if its density matrix
takes the form $\rho = \sum_{i=1}^{\infty} \lambda_i \rho_i^A
\otimes \rho_i^B$, where $0 \le \lambda_i \le 1$, for all $i \in
{N}$, $\sum_{i=1}^{\infty} \lambda_i =1$ and ${\rho}_i^A$
(resp. ${\rho}_i^B$) is a density matrix which involves only $A$ (resp. $B$) coordinates $\widehat{z}^A$ (resp.
$\widehat{z}^B$). The associated Wigner function is:
\begin{equation}
W \rho (z) = \sum_{i=1}^{\infty} \lambda_i W \rho_i^A (z^A) W \rho_i^B (z^B).
\label{eq9}
\end{equation}
and, through the DT, the commutative counterpart in terms of the commutative variables
$\zeta$ is:
\begin{equation}
W \widetilde{\rho} (\zeta) = \sum_{i=1}^{\infty} \lambda_i W \widetilde{\rho}_i^A (\zeta^A) W \widetilde{\rho}_i^B (\zeta^B).
\label{eq10}
\end{equation}
To simplify the manipulation of the above results, let one defines ${\bf \Lambda}$ to be the $2 n\times 2n$ matrix ${\bf \Lambda}= Diag\left[{\bf I^A},\, {\bf \Lambda^B}\right]$, with ${\bf \Lambda^B}= Diag\left[{\bf I},\, -{\bf I}\right]$.
Thus, the transformation
$\zeta \mapsto {\bf \Lambda} \zeta$
amounts to a mirror reflection of Bob's momenta.

Thus, through the PPT criterion, if a Wigner function $W \widetilde{\rho} (\zeta)$ is that of a separable state, then the transformation
\begin{equation}
W \widetilde{\rho} (\zeta) \mapsto W \widetilde{\rho}^{\prime} (\zeta) = W \widetilde{\rho} ({\bf \Lambda} \zeta),
\label{eq11}
\end{equation}
leads to an equivalent  Wigner function.
Hence, if the state $\widetilde\rho$ is separable then
\begin{equation}
\widetilde{\bf \Sigma}^{\prime} + (i/2){\bf J}\geq 0
\label{eq12}
\end{equation}
where $\widetilde{\bf \Sigma^{\prime}}$ is the covariance matrix
of $W \widetilde{\rho}^{\prime} (\zeta)$. 
Eq.~(\ref{eq11}) can then be rewritten in terms of the NC variables as
follows:
\begin{equation}
W {\rho} (z) \mapsto W \rho^{\prime} (z) = W \rho ( {\bf D} z),
\label{eq13}
\end{equation}
where $W \rho (z) = {1 \over \sqrt{\det {\bf \Omega}}} W
\widetilde{\rho} ({\bf S}^{-1} z)$ and $W \rho^\prime (z) = {1
\over \sqrt{\det {\bf \Omega}}} W \widetilde{\rho}^\prime ({\bf
S}^{-1} z)$ are defined accordingly to Eq.~(\ref{eq6}) and ${\bf
D}= {\bf D}^{-1}= {\bf S}{\bf \Lambda}{\bf S}^{-1} = Diag[{\bf
I^A},\,{\bf S^B} {\bf \Lambda^B} ({\bf S^B})^{-1}]$. It follows
from Eq.~(\ref{eq7}) that the covariance matrices ${\bf
\Sigma^{\prime}}$ and ${\bf \widetilde\Sigma^{\prime}}$ of $W
\rho^\prime$ and $W \widetilde\rho^\prime$ are related by ${\bf
\Sigma^{\prime}} = {\bf S}{\bf \widetilde\Sigma^{\prime}} {\bf
S}^T$. Hence, the separability condition Eq.~(\ref{eq12}) can then
be written exclusively in terms of the NC objects
\begin{equation}
{\bf \Sigma^{\prime}} + (i/2){\bf \Omega}\geq 0~. \label{eq13A}
\end{equation}
In addition, notice that ${\bf \Sigma^{\prime}} = {\bf D}{\bf \Sigma} {\bf D}^T$ where ${\bf \Sigma}$ is the covariance matrix of $W\rho$.
Defining ${\bf \Omega^{\prime}} = {\bf D}^{-1} {\bf\Omega} ( {\bf D}^T)^{-1}$ one obtains
\begin{equation}
{\bf \Sigma} + (i/2)   {\bf \Omega^{\prime}}  \ge 0~. \label{eq15}
\end{equation}
We point out that the matrix ${\bf \Omega^{\prime}}$ is simply given by
\begin{equation}
{\bf \Omega}^{\prime} = Diag\left[{\bf
\Omega^A},\,-{\bf \Omega^B}\right],
\label{eq15.1}
\end{equation}
where one has used the definition of ${\bf \Omega}$ and the fact
that ${\bf \Lambda^B}$ is an anti-symplectic transformation, i. e.
${\bf \Lambda^B} {\bf J^B} {\bf \Lambda^B} = - {\bf J^B}$. By
itself, this result is an elegant generalization of the result
from Ref.~\cite{Simon} which states that ${\bf J}$ is replaced by ${\bf
J}^{\prime} = Diag \left[{\bf J}^A, - {\bf J}^B \right]$ under
PPT. We stress the fact that Eq.~(\ref{eq15.1}) is valid assuming
that the DTs take the block-diagonal form ${\bf S}=Diag \left[{\bf
S^A},{\bf S^B}\right]$.

\section{Results for $2$-dim Gaussian states}

A real normalizable phase-space Gaussian function $F(z)$ with covariance matrix ${\bf \Sigma}$ is the NC Wigner function of a quantum separable state if and only if ${\bf \Sigma}$ satisfies the Eqs.~(\ref{eq8}) and (\ref{eq15}) with respect to quantum nature and separability, respectively.
As suggested in \cite{Bernardini13B}, let one consider a Gaussian state written as
\begin{equation}
F(z) = \frac{1}{\pi^4 \sqrt{\det {\bf \Sigma}}} \exp (- z^T {\bf \Sigma}^{-1} z),
\label{eq19}
\end{equation}
which lives on a $8$-dimensional NC phase-space with $n_A = n_B = 2$, ${\bf \Omega^A}= {\bf \Omega^B}$, $\theta_{ij} = \theta\epsilon_{ij}$ and $\eta_{ij} = \eta\epsilon_{ij}$, with $i,\,j = 1,\,2$, and where the covariance matrix is given by,
\begin{equation}
{\bf \Sigma}={b\over 2}\left(
\begin{array}{c c}
{\bf I}_{4}&{\bf \gamma^T}\\
{\bf \gamma}&{\bf I}_{4}
\end{array}\right),~
\end{equation}
for which the DT corresponds to the map ${\bf S}= Diag[{\bf S^A},\,{\bf S^B}]$, with
\begin{equation}
{\bf S^A}= {\bf S^B} = \left(
\begin{array}{c c c c}
\lambda &  0 &  0 & -\theta/2\lambda\\
0 &  \lambda &  \theta/2\lambda & 0\\
0 &  \eta/2\mu & \mu & 0\\
\eta/2\mu& 0 &  0 & \mu
\end{array}\right),
\label{eq21}
\end{equation}
subject to $\lambda\mu = (1 + \sqrt{1 - \eta\theta})/2$, which ensures the map invertibility.
The NC parameters, $\theta$ and $\eta$, are real positive constants satisfying the condition $\theta\eta < 1$.

The NC quantumness and separability criteria are expressed in terms of the so-called NC symplectic spectrum. 
For a $2n \times 2n$ real
symmetric positive-definite matrix ${\bf \widetilde{\Sigma}}$ it is
the set of eigenvalues of the matrix $2 i {\bf J}^{-1} {\bf
\widetilde{\Sigma}}$ and these eigenvalues are called the
Williamson invariants of ${\bf \widetilde{\Sigma}}$. They are all
positive and $\widetilde{\nu}_-$ denotes the smallest
Williamson invariant. By Williamson's Theorem \cite{Simon}, one
can show that ${\bf \widetilde{\Sigma}}$ complies with the RSUP
(${\bf \widetilde{\Sigma}} + (i/2) {\bf J} \ge 0$) if and
only if $\widetilde{\nu}_- \ge 1$. By the same token, one calls the
set of eigenvalues of $2i {\bf \Omega}^{-1} {\bf \Sigma}$ the 
NC symplectic spectrum of ${\bf \Sigma}$ and the eigenvalues are
called the NC Williamson invariants.
The smallest NC Williamson invariant of ${\bf \Sigma}$ is denoted by $\nu_-$.

\begin{figure}
\begin{center}
\vspace{-.3cm}
\includegraphics[scale=.42]{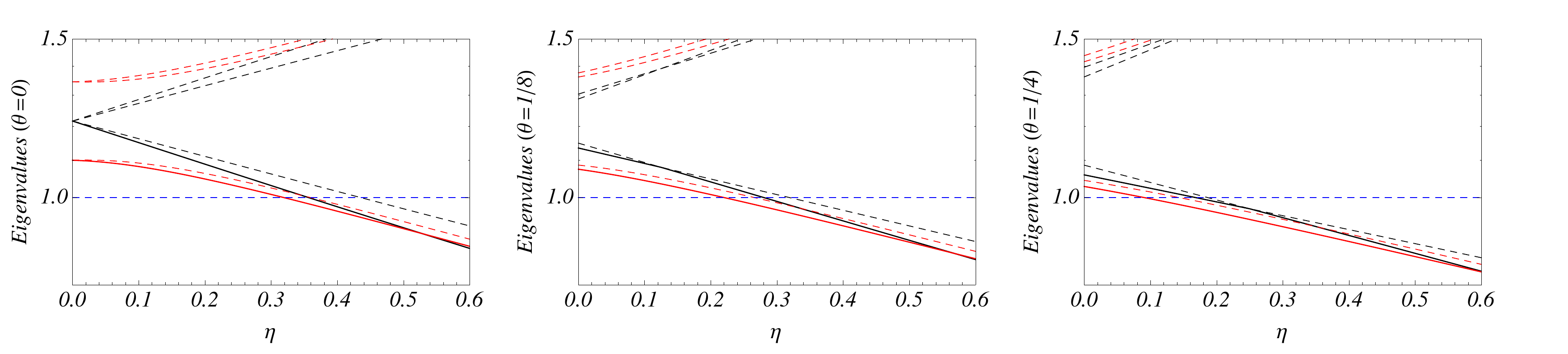}
\includegraphics[scale=.42]{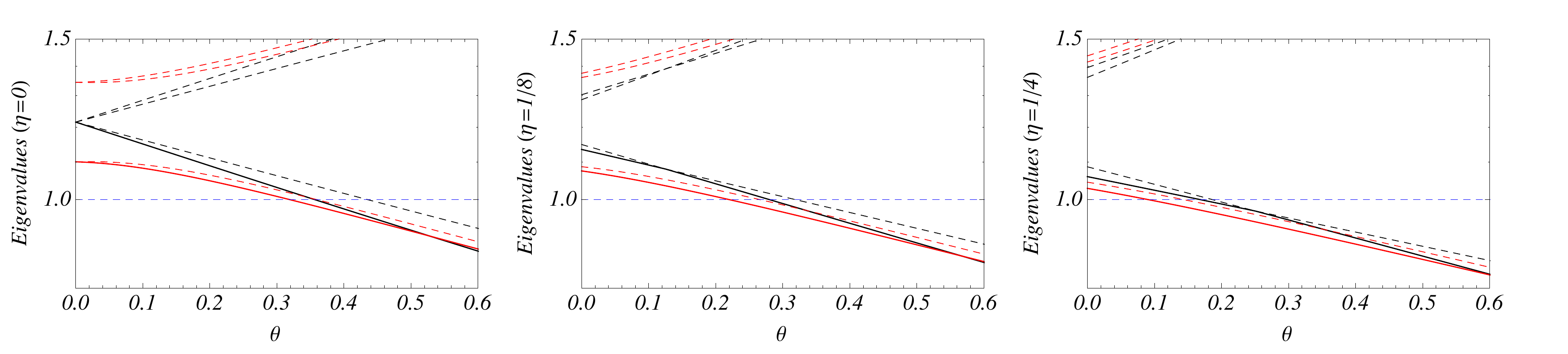}
\vspace{-.2cm}
\caption{Eigenvalues $\nu$ (black lines) and $\nu^\prime$ (red lines) for $\theta = 0,\,1/8$,
and $1/4$ as function of $\eta$ in the range $[0,0.6]$ (first set) and for $\eta = 0,\,1/8$,
and $1/4$ as function of $\theta$ in the range $[0,0.6]$. Solid lines correspond to the respective smallest eigenvalues, $\nu_-$ and $\nu_-^\prime$. Notice that entanglement ($\nu_-^\prime < 1$) coexists with quantum behavior ($\nu_- \geq 1$) for $\eta \neq 0$.
Plots are for $m = R/10$ and $n = 3\sqrt{11}R/10$ and the results are in correspondence with the first example.} \label{res2A}
\end{center}
\end{figure}

\begin{figure}
\begin{center}
\vspace{-.3cm}
\includegraphics[scale = .53]{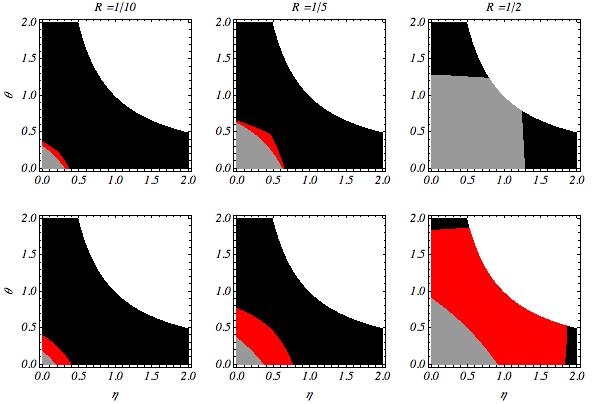}
\vspace{-.2cm}
\caption{The entanglement properties of the NC Gaussian states with the covariance matrix constrained by
$R=1/10,\, 1/5$, and $1/2$. Plots are for $m = R/10$ and $n = 3\sqrt{11}R/10$ (first line) and for $m = 3\sqrt{11}R/10$ and $n = R/10$ (second line). Gray and red regions correspond respectively to separable ($\widetilde{\nu}_- \geq 1$)
and entangled ($\nu_-^{\prime} < 1$) quantum states ($\nu_-^{\prime} \geq 1$). Black region denotes the violation of the
RSUP ($\nu_- < 1$) and the white region is out of the region bound by $\theta \eta \leq 1$.
The results are in correspondence with Fig.~\ref{res2A}.}
\label{res3A}
\end{center}
\end{figure}

{\it First example - } One considers the non-block-diagonal elements of ${\bf \gamma}$ as
\begin{equation}{\bf \gamma}=\left(
\begin{array}{c c c c}
n &  0 &  m & 0\\
0 &  n &  0 & -m\\
m &  0 & -n & 0\\
0 & -m &  0 & -n
\end{array}\right),
\label{eq20}
\end{equation}
\normalsize
where $b,\,m$ and $n$ are real parameters which are assumed, for simplicity, as to be constrained by the relations $R = \sqrt{m^2 + n^2}$, $b = (1 + R)/(1 - R)$.
In this case, the smallest NC Williamson invariants of ${\bf \Sigma}$ and ${\bf \Sigma^{\prime}}$ are given by \cite{Bernardini13B}
\begin{equation}
\nu_- ={1 \over {1-\eta\theta}}{{1+R} \over {1-R}}\sqrt{{{\omega_-} \over 2} -\sqrt{{{\omega_-^2} \over 4} -  \left(1-R^2\right)^2 (1-\eta  \theta )^2}},\nonumber
\end{equation}
\begin{equation}
\nu_-^{\prime} ={1 \over {1-\eta\theta}}{{1+R} \over {1-R}}\sqrt{{{\omega_+} \over 2} -\sqrt{{{\omega_+^2} \over 4} -  \left(1-R^2\right)^2 (1-\eta  \theta )^2}},\nonumber
\end{equation}
respectively, where
$\omega_{\pm} = 2\left(1\pm n^2\right)+\left(1 \mp n^2\right) \left(\eta^2 + \theta^2\right) \pm 2 m^2 (1 + \eta\theta) + n(1 \mp 1)\vert\eta^2 - \theta^2\vert + 2m(1 \pm 1)(\eta + \theta)$,
which allows for examining the role of the NC parameters on the Gaussian entanglement.
The NC quantum nature and the separability of Gaussian states are ensured for $\nu_- \geq 1$ and $\nu_-^{\prime} \geq 1$, respectively.
Entangled quantum states are found in the range  $\nu_- \geq 1 >  \nu_-^{\prime}$.
Setting $\theta$(either $\eta$) equals to constant values, the obtained NC Williamson invariants, $\nu_-$ and $\nu_-^{\prime}$, depend exclusively on $\eta$(or $\theta$), as depicted in Fig. \ref{res2A}.
The standard (commutative) QM limit is obtained by setting $\theta  = \eta = 0$, which implies that $\nu_- = (1 + R)^{3/2}/(1-R)^{1/2}$ and $\nu_-^{\prime} = 1 + R$.
This means that for $0 \le R < 1$, all states are quantum and separable, given that $\nu_-\geq 1$ and $\nu_-^{\prime} \geq 1$, respectively.
Fig.~\ref{res3A} shows how the NC phase-space ($\theta\neq 0 \neq \eta$) induces the entanglement of the corresponding Gaussian states.
The influence of the NC parameters, $\theta$ and $\eta$, on the quantum nature, separability and entanglement of states can be depicted for several values of $R$ (with a degeneracy associated to $m \leftrightarrow n$).

\begin{figure}[h]
\begin{center}
\vspace{-.3cm}
\includegraphics[scale=.4]{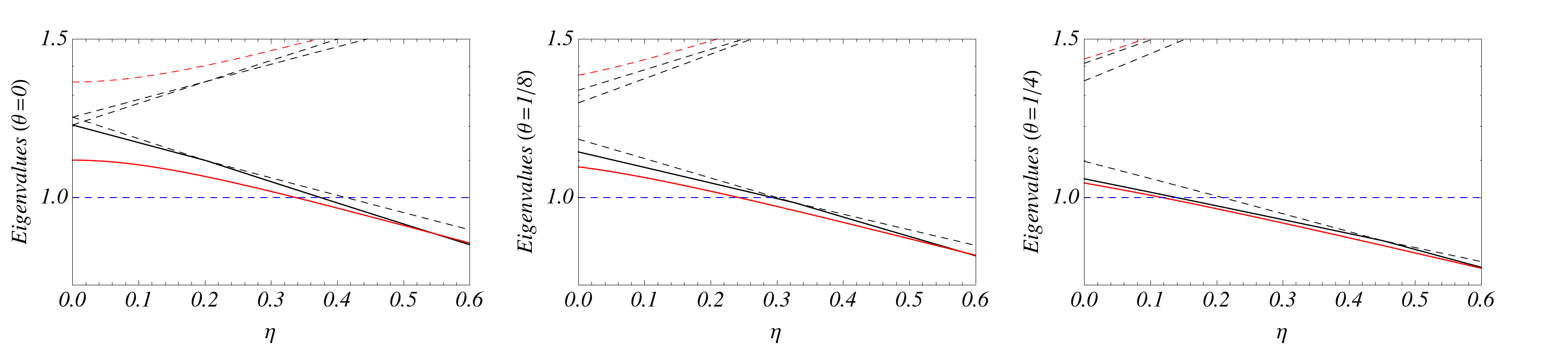}
\includegraphics[scale=.4]{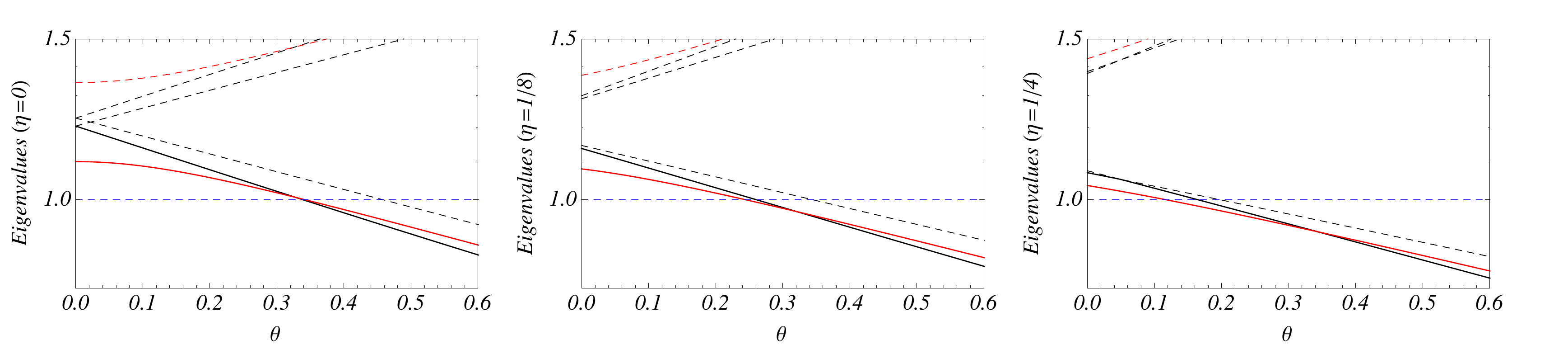}
\vspace{-.2cm}
\caption{Eigenvalues $\nu$ (black lines) and $\nu^\prime$ (red lines) for $\theta = 0,\,1/8$,
and $1/4$ as function of $\eta$ in the range $[0,0.6]$ (first set) and for $\eta = 0,\,1/8$,
and $1/4$ as function of $\theta$ in the range $[0,0.6]$.
Plots are for $m = R/10$ and $n = 3\sqrt{11}R/10$ and the results are in correspondence with the second example.} \label{res2B}
\end{center}
\end{figure}

\begin{figure}[h]
\begin{center}
\vspace{-.3cm}
\includegraphics[scale = .53]{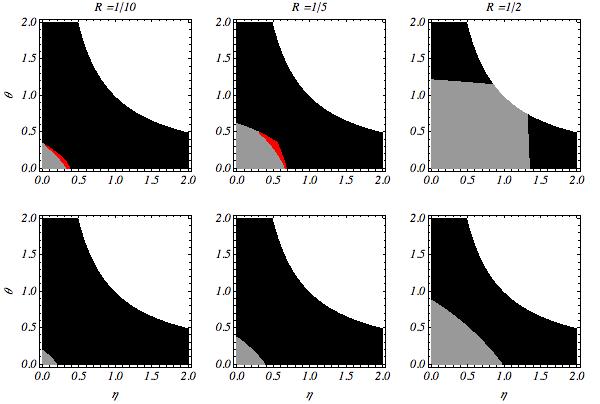}
\vspace{-.3cm}
\caption{The entanglement properties of the NC Gaussian states with the covariance matrix constrained by
$R=1/10,\, 1/5$, and $1/2$. Plots are for $m = R/10$ and $n = 3\sqrt{11}R/10$ (first line) and for $m = 3\sqrt{11}R/10$
and $n = R/10$ (second line). Gray and red regions correspond respectively to separable ($\widetilde{\nu}_- \geq 1$)
and entangled ($\nu_-^{\prime} < 1$) quantum states ($\nu_-^{\prime} \geq 1$). 
The results are in correspondence with Fig.~\ref{res2B}.
Notice that, in this case, the entanglement induced by noncommutativity is highly suppressed, in comparison to the first example}
\label{res3B}
\end{center}
\end{figure}

{\it Second example - } One considers the non-block-diagonal elements of ${\bf \gamma}$ as
\begin{equation}{\bf \gamma}=\left(
\begin{array}{c c c c}
n &  0 &  0 & -m\\
0 &  n &  m & 0\\
0 &  m & -n & 0\\
-m & 0 &  0 & -n
\end{array}\right),
\label{eq20}
\end{equation}
where again the parameters are constrained by the relations $R = \sqrt{m^2 + n^2}$, $b = (1 + R)/(1 - R)$.
In this case, the expression for the smallest NC Williamson invariants are too extensive to be written explicitly.
We just depict the results from Figs.~\ref{res2B} and \ref{res3B} from which one can notice that the symmetry involving the dependence on $\eta \leftrightarrow \theta$ disappears.

\section{Conclusions}

Results derived from an extension of the PPT criterion for investigating the entanglement and the separability in the phase-space NC QM are discussed and the framework has been considered in the analysis of a novel Gaussian state configuration.
Typical constraints on the smallest symplectic eigenvalue of the partially transposed covariance matrix of the state were defined in order to quantify the effect of noncommutativity on the quantumness and separability properties of Gaussian states.
Once again, separable standard QM two-mode Gaussian states have been shown to exhibit quantum entanglement, which is exclusively driven by the NC deformation of the phase-space.
This means that noncommutativity can by itself induce the entanglement of Gaussian states. 
This also reinforces our previous results which imply that
QM effects is a timed version of a more encompassing phase-space NC structure \cite{Bastos,Catarina002,RSUP2}. 

\ack{AEB thanks for the financial support from the Brazilian Agency CAPES (grant nr. AEX 4979/14-1).}

\smallskip
\end{document}